\documentclass[aps,prd,onecolumn]{revtex4-2}

\usepackage{amsmath,amssymb,amsfonts}
\usepackage{graphicx,xcolor}
\usepackage{hyperref}
\usepackage{cleveref}
\crefname{equation}{}{} 
\crefname{figure}{Fig.}{Figs.} 
\crefname{section}{Sec.}{Secs.}
\usepackage[normalem]{ulem}

\newcommand{\mn}{{\mu\nu}}
\newcommand{\ab}{{\alpha\beta}}
\newcommand{\U}{{u}}
\newcommand{\rhoB}{{\rho_B}}
\newcommand{\AS}[1]{{\color{black}{#1}}}

\begin{document}

\title{Close encounters with attractors of the third kind}
\author{Alexander Soloviev}
\affiliation{Faculty of Mathematics and Physics, University of Ljubljana, Jadranska ulica 19, SI-1000, Ljubljana, Slovenia}
\email{alexander.soloviev@fmf.uni-lj.si}
\date{\today}

\begin{abstract}
We report on the existence of a hydrodynamic attractor in the Mueller-Israel-Stewart framework of a fluid living in the novel geometry discovered recently by Grozdanov in \cite{Grozdanov:2025cfx}. This geometry, corresponding to a hyperbolic slicing of dS$_3\times\mathbb{R}$, complements previous analyses of attractors in Bjorken (flat slicing) and Gubser (spherical slicing) flows. The fluid behaves like a sharply localized droplet propagating rapidly along the lightcone. Typical solutions approach the hydrodynamic attractor rapidly at late times despite a Knudsen number exceeding unity, suggesting that the inverse Reynolds number captures hydrodynamization more faithfully since the shear stress vanishes at late times.  
This is in stark contrast to Gubser flow, which has both the Knudsen and inverse Reynolds number becoming small for intermediate times. We close with a comparison to Weyl-transformed Bjorken flow and discuss possible phenomenological applications.
\end{abstract}

\maketitle

\textbf{\textit{Introduction.~}}
Describing the approach of initially non-equilibrium systems to equilibrium is one of the unifying pursuits across a broad range of physical disciplines. In high energy physics, this has been the long time study of how a system thermalizes post collision \cite{Berges:2020fwq}. Thermalization naturally leads to hydrodynamization, how well a system becomes described by hydrodynamics. Interestingly, although hydrodynamics is a theory which is sensible as a derivative expansion, it has been observed that hydrodynamics holds unreasonably well outside of its naive range of applicability, when gradients are large \cite{Romatschke:2017ejr}. 
This tension has been in part resolved via the discovery of hydrodynamic attractors \cite{Heller:2015dha} (see \cite{Soloviev:2021lhs,Jankowski:2023fdz} for reviews), which represent non-equilibrium solutions to which generic evolutions decay. Consequently, the system ``forgets" about its initial configuration as it approaches a hydrodynamic regime.

Two particular flows have been consequential for high energy physics: that of Bjorken flow \cite{PhysRevD.27.140}, describing boost-invariant matter, and radially expanding, boost-invariant Gubser flow \cite{Gubser:2010ze,Gubser:2010ui}.
Attractors have been studied extensively in these backgrounds
\cite{Romatschke:2017vte,Romatschke:2017acs,Behtash:2017wqg,Denicol:2018pak,Behtash:2018moe,Blaizot:2019scw,Kurkela:2019set,Behtash:2019qtk,Behtash:2019txb,Heller:2020anv,Dash:2020zqx,Behtash:2020vqk,Mitra:2020mei,Mitra:2022xtb,Heller:2020hnq,Heller:2021oxl,Chattopadhyay:2021ive,Ambrus:2021sjg,Blaizot:2021cdv,Jaiswal:2022udf,Cartwright:2022hlg,Chen:2022ryi,An:2023yfq,Boguslavski:2023jvg,Nugara:2023eku,Banerjee:2023djb,Mitra:2024zfy,Frasca:2024ege,Wang:2024afv,Spalinski:2025ngd,Chen:2025qao,DeLescluze:2025gaa}, as well as in the case of Yang-Mills kinetic theory \cite{Du:2022bel} and in the classical problem of a gas of 
hard spheres  \cite{Denicol:2019lio}. 
Outside of high energy physics, attractors have recently become the subject of intense interest in the condensed matter community, including ultracold atoms \cite{PhysRevLett.133.173402,Heller:2025yxm}, systems with periodic driving \cite{Mazeliauskas:2025jyi} and superfluid flows \cite{Mitra:2020hbj,Buza:2024jxe}.
Moreover, attractors have begun to be studied in the cosmological context, including with both fixed \cite{Du:2021fok} and dynamical \cite{Buza:2024jxe}
Hubble parameter.

Recently, a new flow has been discovered by Grozdanov \cite{Grozdanov:2025cfx}, exploiting a different foliation of dS$_3$. Whereas the Gubser geometry is a spherical slicing of dS$_3$, the new geometry is instead a hyperbolic slicing (denoted as $\kappa=-1$). Additionally, a flat slicing is possible, which corresponds to Bjorken flow in dS$_3$. Applying the standard arguments of hydrodynamics to this third, hyperbolic geometry leads to a fluid invariant under SO$(1,1)\times $SO$(2,1)_q\times \mathbb{Z}_2$ symmetry. Physically, this symmetry contains boost-invariance, rotations in the transverse directions parameterized by the energy scale $q$, and reflections along the beam axis, respectively.

\begin{figure*}
    \centering
    \includegraphics[width=0.49\linewidth]{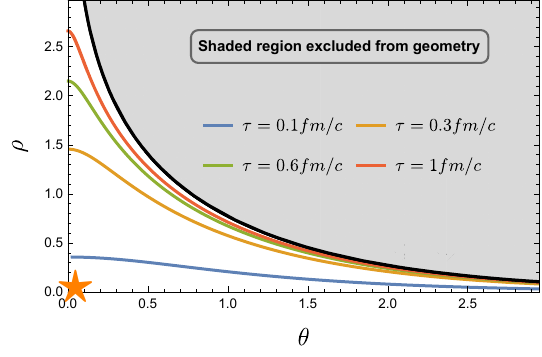}\includegraphics[width=0.49\linewidth,height=0.33\linewidth]{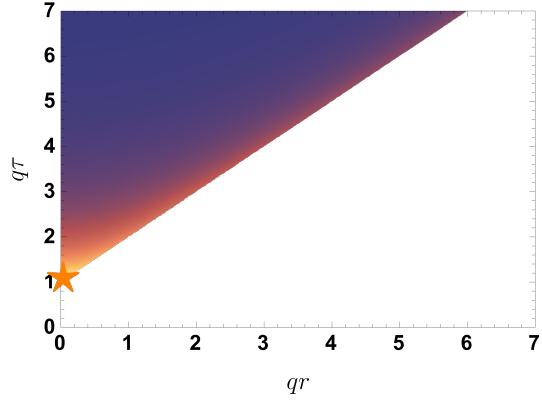}
    \caption{Left: Constant proper time slices viewed in the $(\rho, \theta)$ plane. The lines indicate increasing values of $\tau.$ Right: Temperature profile of the viscous fluid described in text in Milne coordinates subject to the initial condition $\dot\varepsilon(\rho_0=0.1)=2$, with red-blue corresponding to hot-cold. For both plots, the \AS{orange star corresponds to $(q\tau=1,r=0,\phi,\eta)$ in Minkowski space}. }
    \label{fig:coords}
\end{figure*}

In this work, we show for the first time that such a geometry indeed admits an attractor solution in the standard Mueller-Israel-Stewart (MIS) framework \cite{mueller,israel}, \AS{which follows directly from the structure of the MIS equations, where the combination of expansion and dissipative relaxation naturally drives solutions toward a universal trajectory. Our result makes clear how this familiar mechanism appears in this new geometric setting.} Moreover, its dynamics are distinct from both Bjorken and Gubser flow. It is defined in the forward lightcone like Bjorken coordinates, but the \AS{initial proper time of the system is not at}  $\tau=0$, but at $1=q \tau$ at the \AS{radial center, $r=0$}, where $q$ is an associated energy scale of the system at early times. 
When viewed from the flat Milne coordinates, typical solutions for the temperature of the fluid localize along the lightcone and become empty in the central region $r=0$, see \cref{fig:coords}. \AS{In some ways this is reminiscent of the remnant of a wounded nucleus, although in realistic models of the color glass condensate \cite{Gelis:2010nm} such remnants propagate to large longitudinal rapidity, not to large radial distances as is the case here.} Moreover, the Knudsen number, Kn $\sim \tau_\pi \nabla_\mu u^\mu,$ measuring the typical ratio of the microscopic to macroscopic distance, is never vanishingly small in the flow described by Grozdanov. This is very much unlike the situation in Gubser flow, where the shear stress and Kn vanish for intermediate times, which naturally begets a hydrodynamic description. In the present geometry, the inverse Reynolds number vanishes for late times, indicating the hydrodynamic description is valid despite the large value of Kn. Again, this is distinct to Gubser flow where the shear stress freezes to a non-zero constant at late de Sitter times, which corresponds to late time free streaming.

\textbf{\textit{Setup.}}
The metric of interest lives in the $dS_3\times \mathbb{R}$ geometry, with $(\rho, \theta, \phi)$ parameterize the 3 dimensional de Sitter space and is given by \cite{Grozdanov:2025cfx}
\begin{align}\label{eq:metric}
    ds^2 = - d\rho^2 +\sinh^2\rho\, (d\theta^2 + \sinh^2 \theta d\phi^2)+d\eta^2,
\end{align}
where we have taken the de Sitter radius $L=1$, the de Sitter time is  
$0<\rho<\infty$, the range of the angular coordinates is $0\leq\theta<\infty$ and $0\leq\phi<2\pi,$ and $\eta\in \mathbb{R}$ is the rapidity.
Conversion back to flat spacetime in Milne (Bjorken)  coordinates
\begin{align}\label{eq:bjorken}
ds^2_B=-d\tau^2+dr^2+r^2 d\phi^2+\tau^2 d\eta^2,
\end{align}
is given by a coordinate transformation
\begin{align}\label{eq:transform}
    \cosh \rho = \frac{1+q^2(\tau^2-r^2)}{2 q \tau}, \quad \tanh \theta = -\frac{2q r}{1-q^2(\tau^2-r^2)},
\end{align}
followed by the Weyl transform
\begin{align}\label{eq:weyl}
    ds^2_B = \tau^2 ds^2.
\end{align}
The range of de Sitter time leads to the restriction in the Bjorken coordinates
\begin{align}\label{eq:restriction}
  \AS{ \frac{1}{q}<\tau-r,}
\end{align}
i.e.~that $q$ sets the initial energy scale. To better orient the reader \AS{and neglecting the azimuthal and rapidity coordinates}, we note that the spacetime begins in the Milne coordinates at $(\tau,r)=(q^{-1},0)$, which in the new geometry this corresponds to $(\rho,\theta)=(0,0)$.

We now develop the theory of a conformal viscous fluid in the background \cref{eq:metric}. The energy momentum tensor is given by
\begin{align}
    T^\mn = (\varepsilon+p)\U^\mu \U^\nu+ p g^\mn +\Pi^\mn,
\end{align}
where $\varepsilon$ is the energy density, $p$ is the pressure and $\Pi^\mn$ is the dissipative tensor. We will assume that the four velocity is at rest in this metric, $\U^\mu=(1,0,0,0)$.  The dissipative tensor is orthogonal to the four velocity, $\Pi^\mn \U_\mu =0,$ and traceless, $\Pi^\mn g_\mn=0.$ \AS{Furthermore, we will work with conformal fluids, where $T^\mu_{\mu}=0$.}

The evolution equations will be given by the conservation
\begin{align}\label{eq:conservation}
    \nabla_\mu T^\mn=0,
\end{align}
and the Mueller-Israel-Stewart \cite{mueller,israel,Baier:2007ix} equation, describing the relaxation of the dissipative tensor
\begin{align}\label{eq:mis}
   \left(\tau_\pi u^\mu \nabla_\mu+1\right) \Pi^\ab = - \eta \sigma^\ab,
\end{align}
where $\tau_\pi$ is the relaxation time, $\eta$ is the shear viscosity and $\sigma^\mn=\Delta^{\mu\alpha}\Delta^{\nu\beta} \left(\nabla_\alpha \U_\beta +\nabla_\beta \U_\alpha\right)-\frac{2}{3}\Delta^{\mn}\nabla_\alpha \U^\alpha$ is the symmetric, transverse and traceless shear tensor, where the spatial projection $\Delta^\mn=\U^\mu \U^\nu+g^\mn,$ see e.g.~\cite{Baier:2007ix}. 
We will parameterize the dissipation tensor as follows
\begin{align}
\Pi^\alpha_{\phantom{\alpha}\beta}=\text{diag}\left(0,\frac{\Pi}{2},\frac{\Pi}{2},-\Pi\right).
\end{align}

Taking the energy density, pressure and the components of the dissipative tensor to be functions of $\rho$ only, we see that \cref{eq:conservation,eq:mis} in the background \cref{eq:metric} are 
\begin{subequations}\label{eq:tobesolved}
\begin{align}  
\dot\varepsilon+(\varepsilon+p)(2-\pi)\coth{\rho}
&=0,\\
    \tau_\pi\dot \pi +\pi+\left(\frac{4}{3} \tau_\pi \pi^2-\frac{\eta}{\varepsilon}\right)\coth \rho&=0,
\end{align}
\end{subequations}
where we introduced the dimensionless quantity $\pi\equiv-\Pi/(\varepsilon+p)$. We see immediately that in the inviscid case ($\pi=0)$, the energy density for the conformal fluid is given by \cite{Grozdanov:2025cfx}
\begin{align}
    \varepsilon(\rho)=\varepsilon_0 \sinh^{-8/3}\rho,
\end{align}
where we used that the conformal fluid equation of state is $p=\varepsilon/3.$
To transform the energy density back to Bjorken coordinates, one uses the coordinate transformation \cref{eq:transform} followed by the Weyl rescaling, to find $\varepsilon(\tau)=\tau^{-4}\varepsilon(\rho)\sim \tau^{-4/3}$ \AS{in the limit where the transverse extent goes to zero, $q\rightarrow 0.$}

 \AS{For a conformal fluid t}he only scale remaining is the temperature, $\varepsilon\sim T^4.$ Transport coefficients will be expressed via dimensionless quantities
\begin{align}
    \tau_\pi = C_\tau \varepsilon^{-1/4}, \quad \eta = C_\eta s,
\end{align}
where $s=(\varepsilon+P)/T$ is the entropy density.
For specificity, we will choose to work with dimensionless transport coefficients computed from holography \cite{Bhattacharyya:2007vjd}
\begin{align}
    C_\eta = \frac{1}{4\pi}, \quad C_\tau = \frac{2-\log 2}{2\pi}.
\end{align}

For future convenience, if we choose to instead parameterize the shear viscosity in terms of the relaxation time $\eta=(\varepsilon+p)\tau_R/5$ \AS{as computed in, e.g.~RTA kinetic theory \cite{Romatschke:2015gic}}, the equations \cref{eq:tobesolved} are then
\begin{subequations}\label{eq:theeqs}
\begin{align}\label{eq:cons2}
    \frac{1}{T}\frac{d T}{d\rho}+\frac{2}{3} \coth \rho-\frac{1}{3} \pi  \coth \rho &=0,\\
    \frac{d\pi}{d\rho}+\frac{\pi }{\tau_R}+\frac{4}{3} \pi^2 \coth \rho
   &=\frac{4 }{15} \coth \rho .\label{eq:mis2}
\end{align}
   \end{subequations}
Comparing to eqs.~14-15 of \cite{Denicol:2018pak}, we see that the form of the equations is very similar. However, we reiterate that the coordinate range is quite different, as in the Gubser case\footnote{Throughout the text the de Sitter time in the hyperbolic slicing will be referred to as $\rho$, while in the flat and spherical slicing it will be denoted as $\rho_B$ and $\rho_G$, respectively. The same will hold for other variables whenever confusion could occur.} time stretches for all $\rho_G\in \mathbb{R}.$

\begin{figure*}
    \centering
    \includegraphics[width=0.49\linewidth]{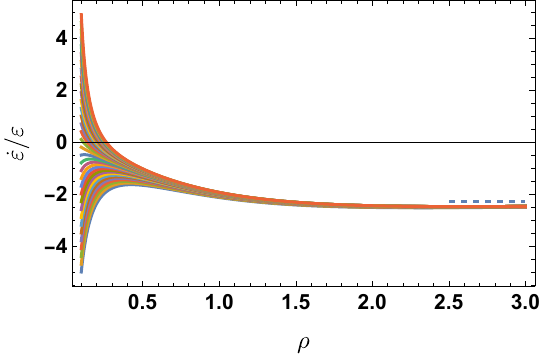}
    \includegraphics[width=0.49\linewidth]{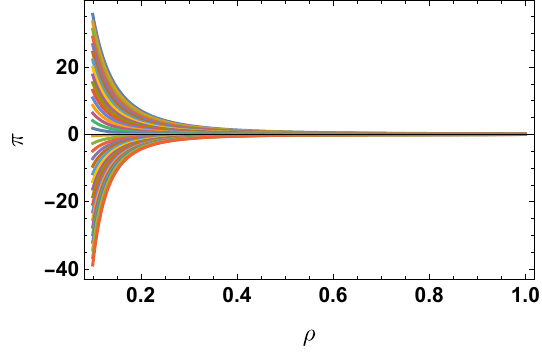}
    \caption{Left: Arbitrary initial conditions approach the same universal curve, for fixed initial energy density, $\varepsilon_0=1,$ with varying initial time variation of the energy density, $\dot\varepsilon_0.$ The late time behavior \cref{eq:latetime} is shown as a dashed line. Right: Dimensionless shear stress tensor for the same initial conditions.}
    \label{fig:attractor}
\end{figure*}

\textbf{\textit{The hydrodynamic attractor in hyperbolic dS$_3\times\mathbb{R}$.}}
To better characterize the attractor, we introduce the dimensionless combination\footnote{Recall that we have taken $L=1$ and that the derivative is with respect to the dimensionless combination $\rho/L$.} 
\begin{align}
    f=\frac{\dot\varepsilon}{\varepsilon}.
\end{align}
We numerically solve \eqref{eq:tobesolved} and provide a characteristic plot of a variety of initial conditions in  \cref{fig:attractor}, initializing at $\rho_0=0.1$ for the same fixed initial energy density, $\varepsilon_0=\varepsilon(\rho_0)$, while varying its time derivative $\dot\varepsilon_0=\dot\varepsilon(\rho_0).$ It is easy to see that by the time $\rho\approx 1,$ the curves collapse to the same curve. At late times $\rho\rightarrow\infty$, $\dot f$ and $\varepsilon$ vanish, leading to a quadratic equation
\begin{align}
   9 C_\tau f^2   +
   48 C_\tau f+64
   C_\tau-16 C_\eta =0,
\end{align}
from which we analytically determine that the physical value the solutions approach is
\begin{align}\label{eq:latetime}
    f_\infty=\frac{4 \left(\sqrt{C_\eta}-2 \sqrt{C_\tau}\right)}{3 \sqrt{C_\tau}}.
\end{align}
Note that unlike in the Bjorken or Gubser flow attractors, the new attractor is not monotonic, approaching \cref{eq:latetime} from below. 
Curiously, the other possible solution at late times, corresponding to the repulsor in this geometry, is given by
\begin{align}
    \tilde f_\infty =-4\frac{ 2 \sqrt{C_\tau}+\sqrt{C_\eta}}{3 \sqrt{C_\tau}}.
\end{align}
This is proportional to the value the Bjorken attractor solution takes for small $w=\tau T$, cf. eq.~(12) in \cite{Heller:2015dha}. Indeed, the unstable early time solution of the Bjorken solution of \cite{Heller:2015dha} corresponds to the stable late time value the system evolves for the present solution in the new geometry \cref{eq:metric}.

We now turn our attention to understanding when hydrodynamics is valid. A straightforward measure is the Knudsen number, which is the ratio of a microscopic to a macroscopic (hydrodynamic) length scale. In the present case, following the notation in \cite{Chattopadhyay:2019jqj},
\begin{align}
    \text{Kn}\sim\frac{l}{L} \sim \tau_\pi \nabla_\mu u^\mu \sim \varepsilon(\rho)^{-1/4}\coth \rho,
\end{align}
which is expressly divergent for large $\rho$. An alternative measure of the viability of hydrodynamics is the inverse Reynolds number, measuring the relative size of dissipative forces
\begin{align}
    \text{Re}^{-1}\sim \sqrt{\pi^\mn \pi_\mn},
\end{align}
which we see from the right panel of \cref{fig:attractor} vanishes for large $\rho.$ Likewise, considering the entropy production to second order \cite{Baier:2007ix},
\begin{align}
    \nabla_\mu( s u^\mu)= \frac{\eta}{2T}\sigma^\mn \sigma_\mn+ \mathcal{O}(\partial^3),
\end{align}
we see that the system progressively produces less entropy in the limit $\rho\rightarrow\infty,$ which indicates that the system is approaching the inviscid hydrodynamic regime at late times.

\begin{figure}[b]
    \centering    \includegraphics[width=0.55\linewidth]{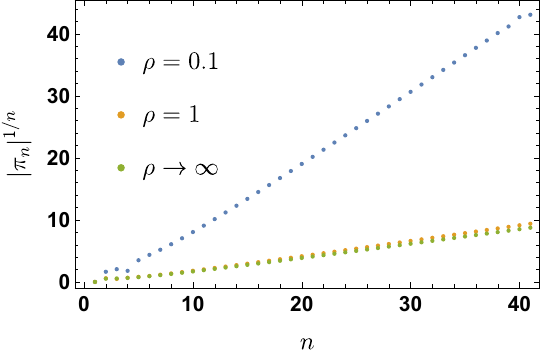}
    \caption{Large order behavior of the hydrodynamic series \cref{eq:expansion} diverges factorially for all $\rho$.}
    \label{fig:hydro-coeff}
\end{figure}

Following the discussion in \cite{Denicol:2017lxn,Denicol:2018pak} for Gubser flow, we convert \eqref{eq:mis2} to a perturbative problem by expanding the shear stress in powers of the relaxation time
\begin{align}
    \pi &= \sum_{n=0}^\infty \pi_n(\rho) \tau_R^n. \label{eq:expansion}
\end{align}
In \cite{Denicol:2018pak}, the authors note that the derivative of $\pi^\mn$ with respect to $\rho$ will also contain $\tau_R\sim 1/T$ dependence. Using the chain rule and \cref{eq:cons2}, we determine the following equation of motion for $\pi_n$
\begin{align}
&\sum_{n=0}^\infty\tau_R^n\left[\tau_R\frac{d\pi_n}{d\rho}+\left(1+\frac{2 n\coth\rho}{3}  \tau_R\right)\pi_n\right]\nonumber\\
&+\frac{\coth\rho}{3}\sum_{n,m=0}^\infty (4-n)\pi_n \pi_m \tau_R^{n+m+1}
   =\frac{4 \tau_R }{15} \coth \rho .
\end{align}
It is straightforward to determine the solution order by order in the relaxation time, with the lowest orders given by
\begin{align}    
    \pi&=\frac{4}{15}\tau_R\coth \rho -\frac{4}{15}\tau_R^2\left(1-\frac{1}{3} \coth^{2}\rho\right) +\mathcal{O}(\tau_R^3),
\end{align}
which we show for various $\rho$ to higher order in \cref{fig:hydro-coeff}. Clearly, the terms are growing factorially, indicating that the gradient expansion is divergent with zero radius of convergence and in need of treatment via analytic techniques like Borel resummation \cite{Heller:2015dha,Romatschke:2017acs,Grozdanov:2019kge,Grozdanov:2019uhi,Aniceto:2024pyc}. 
It is fruitful to contrast the expansion with the Gubser flow case \cite{Denicol:2018pak}, where all odd terms vanish when $\rho_G=0$ (corresponding to precisely when $\pi$ vanishes, Kn is small and the gradient expansion is sensible even if divergent). Here, there is no value of $\rho$ for which all even/odd order terms drop out. Moreover, unlike in the Gubser flow case where the change in $\rho_G$ barely modified the results, \cref{fig:hydro-coeff} shows appreciable differences in slope as $\rho\rightarrow0$.

We now turn to interpreting the solutions by comparing it to the situation in standard and uplifted Bjorken flow. We begin by inverting \cref{eq:transform} to find
\begin{align}
    q \tau = \frac{\text{csch}\, \rho}{\coth \rho-\cosh\theta}, \quad qr=\frac{\sinh\theta}{\coth \rho-\cosh\theta}.
\end{align}
From here it is straightforward to provide an overview of the collision\AS{, focusing on the $(\tau,r)$ coordinates with the understanding that the azimuthal and rapidity coordinates do not play a leading role}. The origin of the future directed lightcone is set at $q \tau_0=1$ and $r_0=0,$ which corresponds to the origin in $(\theta,\rho)$ coordinates. For subsequent proper time, the maximum extent is limited \AS{by \cref{eq:restriction}}. At late times and/or large transverse extent, the geometry ends at the curve paramterized by $\coth\rho=\cosh\theta.$ We show this in \cref{fig:coords} for $q=1/\tau=1/(0.07\text{fm/c})$, which corresponds to the characteristic timescale for the total overlap of colliding nuclei, following \cite{Gubser:2010ui}. 
Further comparison can be made to Bjorken flow by transforming the solution of \cref{eq:theeqs} to the Milne coordinates $(\tau,r)$ via \eqref{eq:transform} and the Weyl transform \cref{eq:weyl}. We show a characteristic plot in the right panel of \cref{fig:coords}. The temperature profile drops rapidly in the \AS{radial} center of the coordinates and localizes primarily along the lightcone.

\begin{figure*}
    \centering    \includegraphics[width=0.49\linewidth]{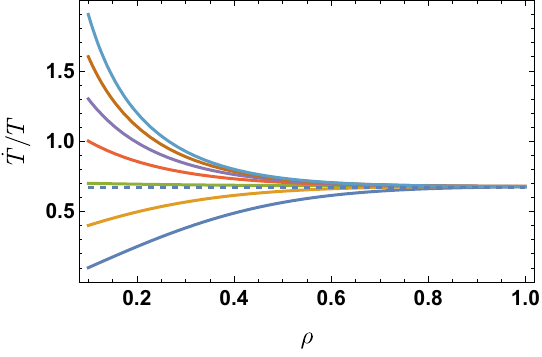}
\includegraphics[width=0.49\linewidth]{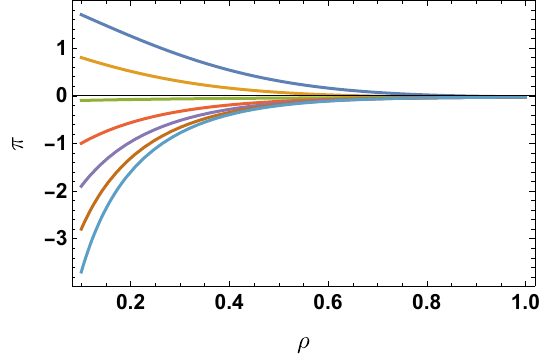}
    \caption{Attractor of the flat foliation of dS$_3\times \mathbb{R}$ defined by the metric \cref{eq:flat-metric}. Left: the time derivative of the temperature \AS{normalized by the temperature} tends to $\frac{2}{3}$ in this space. Right: at the same time, the dissipation drops to zero.}
    \label{fig:flat-attractor}
\end{figure*}

Now, we will opt to stay in the dS$_3\times \mathbb{R}$ spacetime and consider the flat foliation ($\kappa=0$) as described in \cite{Grozdanov:2025cfx}. This will lead to an apples to apples comparison. Note that in this case, the geometry is not Ricci flat, with physical observables like the temperature needing a Weyl rescaling to recover results from Bjorken flow as formulated in $\mathbb{R}^{3,1}.$
The metric in this case reads \cite{Grozdanov:2025cfx}
\begin{align}\label{eq:flat-metric}
    ds^2_{\rm flat}=-d\rhoB^2+e^{-2\rhoB}\left(d\theta_B^2+\theta_B^2 d\phi_B^2\right)+d\eta_B^2,
\end{align}
where $\infty>\rhoB>-\infty$ and the other coordinates have the same domain as written below \cref{eq:metric}.
Completely analogous to the hyperbolic case discussed above, we can solve \cref{eq:conservation,eq:mis} in the above metric, which corresponds to the Bjorken flow in the uplifted dS$_3\times\mathbb{R}$ space. For the same choice of parameters that led to \cref{eq:theeqs}, the particularly simple equations are given by
\begin{subequations}
\begin{align}\label{eq:flat-cons}
    \frac{1}{T}\frac{d T}{d \rhoB }+\frac{\pi}{3}-\frac{2}{3}&=0,\\
   \frac{d\pi}{d\rhoB}+\frac{\pi}{\tau_R}-\frac{4}{3} \pi^2+\frac{4}{15}&=0.\label{eq:flat-mis}
\end{align}    
\end{subequations}
The logarithmic derivative of the temperature and the shear stress are shown for a variety of initial conditions in \cref{fig:flat-attractor}. Like in \cite{Heller:2015dha}, the solutions decay in a way that $\pi$ rapidly dies out, while the logarithmic derivative of the temperature tends to $2/3$ for late times. Recalling that $\rhoB=\ln \tau$ to transform back to Bjorken time \cite{Grozdanov:2025cfx}, we see that at late times, the temperature is given by $T(\rhoB)\sim e^{2\rhoB/3}\sim \tau^{2/3}$, which following a Weyl transform leads to the usual late time temperature dependence of Bjorken flow, $T(\tau)=T(\rhoB)/\tau\sim \tau^{-1/3}$. Note that here, the Knudsen number is $\sim \tau_\pi\nabla_\mu u^\mu = -2 C_\tau/T(\rho_B)$. 
Thus, in contrast to the previous geometry, we see that both the Knudsen number and the inverse Reynolds number become small at late de Sitter times.

As a curious aside, in the case $\tau_R$ is a constant, \cref{eq:flat-mis} has a neat closed form solution given by 
\begin{align}\label{eq:temp-uplift}
    T(\rhoB)=e^{ (\frac{2}{3}-\frac{1}{8\tau_R})\rhoB}\cosh^{1/4}\left(\sqrt{1+\frac{64\tau_R^2}{45}}\frac{\rhoB+c_0}{2\tau_R}\right),
\end{align} where $c_0$ is an integration constant. The expression for $\pi(\rhoB)$ follows immediately from \cref{eq:flat-cons} and corresponds to a hyperbolic tangent. Solving the same system of equations for constant relaxation time in the usual Bjorken flow \cref{eq:bjorken} leads to a much more complicated expression. 

Although a constant relaxation time explicitly breaks conformality by introducing another scale, this may be more than a mathematical curiosity. For the corresponding constant temperature scale, $T_R \sim 1/\tau_R$, we should consider the dimensionless ratio of the temperatures $T_R/T(\rho_B)$. For sufficiently high temperatures,  $T_R/T(\rho_B) \ll 1$, conformality is approximately restored and \cref{eq:temp-uplift} is a valid approximate solution. This is analogous to QCD, which becomes approximately conformal when the temperature is much higher than the QCD scale, $T\gg \Lambda_{\rm QCD}$.

\textbf{\textit{Discussion.~}}
In this work, we provided the first description of a conformal fluid in the hyperbolic slicing of dS$_3\times \mathbb{R}$ \cite{Grozdanov:2025cfx} evolving in the  MIS framework. In this geometry, we saw that the fluid moves ultrarelativistically as a localized droplet. As a natural consequence of the geometry, the resulting solutions approached an attractor solution rapidly in de Sitter time. Intriguingly, and quite different to attractors found in Bjorken/Gubser flow, the approach to the attractor was found to be generically non-monotonic. Curiously, the Knudsen number is never vanishingly small 
in the dS$_3\times \mathbb{R}$ coordinates. Nonetheless, the solutions rapidly decay towards a universal attractor curve. The shear stress decays at a similar rate, indicating that the inverse Reynolds number is perhaps a better measure of the approach to the attractor in this case. 

To better understand the solution, a comparison was made to the Bjorken flow in the Weyl transformed dS$_3\times\mathbb{R}$ space. We saw that the solutions in this space similarly rapidly lose their knowledge of initial conditions. This is due to the vanishing $\pi$, notwithstanding the large constant Knudsen number. Exploring the flat geometry in the uplifted space bore an interesting novel side result: a particularly simple analytic solution \cref{eq:temp-uplift} of the MIS equations in Bjorken flow in the uplifted geometry for constant relaxation time. This solution holds in the high temperature regime, when the system can be argued to be approximately conformal.

A natural next step is to study the consequences of such a geometry in microscopic theories. This would include adapting the discussion to weakly coupled kinetic theory, studying how the distribution function evolves in such a background. It may be possible to determine an exact solution to the Boltzmann equation in a simplified setting, such as the relaxation time approximation, as has been done previously in both Bjorken \cite{Florkowski:2014sfa} and Gubser \cite{Denicol:2014xca} coordinates. Work in this direction may shed further light on the nonthermal fixed point \cite{Berges:2008wm,Berges:2014bba,PhysRevLett.114.061601}, a far-from-equilibrium feature found in a wide range of systems, from cold atomic gases to heavy ion collisions. Similarly, strongly coupled dual descriptions based on the holographic principle have been worked out for Bjorken \cite{Janik:2005zt,Janik:2006ft} and recently for Gubser \cite{Banerjee:2023djb,Mitra:2024zfy} flows. The corresponding holographic construction with the boundary metric given by \cref{eq:metric} remains to be determined. These microscopic theories will provide a useful laboratory, for example as an arena to apply analytic techniques relevant to hydrodynamic attractors, such as Borel resummation \cite{Aniceto:2024pyc,Spalinski:2025ngd} or to determine correlation functions \cite{Kamata:2020mka,Banerjee:2022aub}, as well as eventual phenomenological exploitation.

\acknowledgments

I would like to thank Sa\v so Grozdanov, Toshali Mitra, Ayan Mukhopadhyay and Rajeev Singh for helpful discussions. 
 I am supported by funding from the Horizon Europe
research and innovation programme under the Marie Skłodowska-Curie grant agreement
No. 101103006, the project N1-0245 of Slovenian Research Agency (ARIS) and financial
support through the VIP project UNLOCK under contract no. SN-ZRD/22-27/510. I also thank ECT* for support at the workshop ``Attractors and thermalization in nuclear collisions and cold quantum gases", where I benefited from useful discussions. \\
\textbf{Data availability.}
The data that support the findings of this article are openly available  \cite{data}.

\bibliographystyle{apsrev4-2}
\bibliography{Main}

\end{document}